\documentstyle[11pt,aaspp4]{article}

\def\ir08{\hbox{IRAS 08572+3915}}
\def\mum{\hbox{\micron~}}
\def\lir{{$L_{\rm IR}$}}
\def\lco{$L_{\rm CO}$}

\def\lsun{$L_\odot$}
\def\msun{$M_\odot$}
\def\kms{\hbox{km s$^{-1}$}}
\def\trademark{{\ooalign{\hfil\raise.07ex\hbox{\tiny
R}\hfil\crcr\mathhexbox20D}}}
\begin{document}
\title{The Deep Silicate Absorption Feature in \ir08 \\ and Other
Infrared Galaxies} 
\author{C.~C.~Dudley\altaffilmark{1} \& C.~G.~Wynn-Williams\altaffilmark{1,2}} 
\affil{Institute
for Astronomy, University of Hawaii \\2680 Woodlawn Dr., Honolulu, HI
96822} \altaffiltext{1}{Visiting Astronomer at the United Kingdom
Infrared Telescope, which is operated by the Joint Astronomy
Centre on behalf of the U.K.~Particle Physics and Astronomy Research
Council.}
\altaffiltext{2}{Visiting Astronomer at the NASA Infrared
Telescope Facility, which is operated by the University of Hawaii
under contract to NASA.}  
\authoraddr{Institute for
Astronomy\\University of Hawaii\\2680 Woodlawn Dr. \\ Honolulu, HI
96822}

\begin{abstract}

New mid-infrared (10 and 20 \micron) spectro-photometry of the
ultraluminous infrared galaxy \ir08 is presented.  The 10 \mum
spectrum reveals a deep silicate absorption feature, while the
20 \mum spectrum shows no clear evidence for an 18 \mum
silicate absorption feature.  An interstellar extinction curve
is fitted to \ir08 and two other deep silicate infrared
galaxies, NGC 4418 and Arp 220.  It is found that pure
extinction cannot explain the spectral energy distributions of
these sources.  On the other hand, both the strength of the
silicate absorption and the overall spectral energy
distributions of the three galaxies agree well with scaled-up
models of galactic protostars.  From this agreement, we
conclude that the infrared emission comes from an optically
thick dust shell surrounding a compact power source.  The size
of the power source is constrained to be smaller than a few
parsecs.  We argue that a significant portion of the total
luminosities of these galaxies arises from an active galactic
nucleus deeply embedded in dust.

\end{abstract}
\keywords{dust --- galaxies: individual: \ir08 --- NGC 4418 --- Arp 220 --- 
galaxies: active --- galaxies: nuclei --- infrared: galaxies}

\vskip1in

Preprint IfA-97-31 aka astro-ph/9705242:  To Appear in The Astrophysical Journal.  

\newpage

\section{Introduction}

The nature of the power source in high-luminosity infrared
galaxies is still uncertain even though the existence of
these galaxies was first suggested (e.g.,~Kleinmann \&
Low~\markcite{sil-kl70}1970) more than 25 years ago.  While the
nature of the infrared emission from these galaxies is almost
universally thought to be due to the reprocessing of
shorter-wavelength radiation through absorption and
reemission by dust, this averaging process leaves few clues
as to the nature of the original source of shorter-wavelength
radiation.  Two power sources have been shown to be important
through observations at many wavelengths: starbursts (Gehrz,
Sramek, \& Weedman~\markcite{sil-ge83}1983; Joseph \&
Wright~\markcite{sil-jo85}1985) and buried active galactic
nuclei (AGNs) (Sanders et al.~\markcite{sil-sa88}1988).  Among
galaxies with luminosities in the range $10^{11}$--$10^{12}$
\lsun, massive stars created in a burst of star formation are
the main power source, but there is evidence that at
luminosities above $10^{12}$ \lsun, AGNs may be an
increasingly important power source (Veilleux et
al.~\markcite{sil-ve95}1995; Lonsdale, Smith, \& Lonsdale~\markcite{lo93}1993).

Dust obscuration can hide the optical and near-infrared
spectral signatures that might indicate the presence of an
AGN, but dust emission provides clues that can reveal the
dominant power source.  By examining dust emission between 8
and 13 \mum in 60 galaxies, Roche et al.~\markcite{sil-ro91}(1991)
have shown that the 8--13 \mum spectra of galaxies can be
classified into three types: (1) Starbursts and other galaxies
with \ion{H}{2} regionlike optical spectra show 8--13 \mum
spectra dominated by a family of infrared bands commonly
attributed to polycyclic aromatic hydrocarbon (PAH) molecules
(Puget \& L\'eger~\markcite{sil-pu89}1989, but see also
Duley~\markcite{sil-du89}1989; Sakata \& Wada~\markcite{sil-sak89}
1989; and Ellis et al.~\markcite{sil-el94}1994 for other
laboratory analogues).  PAH emission thus betrays the presence
of hot stars as an important power source for the overall
infrared emission.  (2) Quasars and Seyfert 1 galaxies often
show flat featureless spectra between 8 and 13 \micron. (3) A
minority of galaxies, most of which have Seyfert 2 optical
spectra, show a silicate absorption feature.  Thus,
energetically dominant AGNs have 8--13 \mum spectra that are
quite different from starbursts.  Distinguishing these three
types of 8--13 \mum spectra requires only moderate spectral
resolution.

In this paper, we present new observations of the
ultraluminous infrared galaxy \ir08 ($L_{[1-1000
\micron]}=2\times 10^{12}$ \lsun, $d$ = 233 Mpc, $H_0$ = 75 \kms
Mpc$^{-1}$) that show it to have an extremely deep silicate
absorption feature centered near 9.7 \micron.  \ir08 has only
the third known example of a deep silicate feature among all
observed galaxies with $L_{[1-1000 \micron]} > 10^{11}$
\lsun; the others are NGC 4418 and Arp 220 (Roche et
al.~\markcite{sil-ro86}1986; Smith, Aitken, \&
Roche~\markcite{sil-sm89}1989).  It is our thesis here that the
presence of a deep silicate feature in these three galaxies
cannot be explained by a simple cold dust screen model, but
that these three galaxies contain a power source surrounded
by a optically thick shell of emitting dust with a geometry
that resembles that of a scaled-up model protostar.  Such a
scaling (or indeed simply the notion that the dust emission
is optically thick) leads to the conclusion that a major power
source of deep silicate infrared galaxies is contained within
a region too small to be attributed to a starburst.  An
energetically dominant AGN is the most likely alternative.

\section{Observations and Data Reduction}

\subsection{Spectroscopy}

IRAS 08572+3915 was observed on the nights of UT 1992
February 16 and UT 1994 February 8 at 8--13 \micron, and on
the night of UT 1994 February 7 at 17--24 \micron.  The
cooled grating spectrometer (CGS3) (see Cohen \&
Davies~\markcite{sil-coh95}1995 for a description) mounted at
the Cassegrain focus of the 4 m United Kingdom Infrared
Telescope (UKIRT) on Mauna Kea in Hawaii was used to make the
observations.  The data of UT 1992 February 16 and UT 1994
February 7 were obtained under good conditions. The 8--13
\mum data of UT 1994 February 8 were of poor quality due to
both variable transparency and thermal emission from cirrus
clouds, and will not be discussed further except to say that
they tend to confirm the data taken under good conditions,
albeit at a much lower signal-to-noise ratio.

Observations at 8--13 \mum were performed using the
low-resolution grating mode of CGS3 with a resolution of
$\lambda/\Delta\lambda\approx50$ and a circular aperture with
5\farcs 5 full width at 10\% power, while observations at
17--24 \mum used the 20 \mum grating mode with
$\lambda/\Delta\lambda\approx 70$ and a smaller 3\farcs 26
aperture to improve the signal-to-noise ratio by admitting
less background emission.  In both cases observations were
centered on the 3.6 cm radio position published by Condon et
al.~\markcite{sil-co91}(1991). The centers of the apertures
were determined with respect to the optical guide camera by
peaking up on infrared bright stars, and the apertures were
centered on the radio position by offsetting from a nearby
Carlsberg Meridian Catalog star.  Pointing was maintained to
\hbox{$\sim 1$\arcsec} rms by auto-guiding on a field star in
the guide camera field of view.

Two interlacing grating positions were required to sample fully
the spectral resolution, resulting in spectra of 64 data points
spaced by 0.1 \mum for 8--13 \mum and by 0.14 \mum for 17--24
\micron.  For the 8--13 \mum spectrum the observations consist
of 60 beam-switched pairs of 30 s each (on source) and for the
17--24 \mum spectrum, 96 pairs of 20 s each, for each grating
position. Grating shifts occurred after 10 pairs at 8--13 \mum
and after 16 pairs at 17--24 \micron.  Background subtraction
was performed with a beam separation of \hbox{30\arcsec} for
the 8--13 \mum data and \hbox{20\arcsec} for the 17--24 \mum
data with the chop direction NE--SW in both cases.

For the 8--13 \mum spectrum, data shortward of 7.7 \micron,
between 9.2 and 10.1 \micron, and longward of 13.2 \mum have
been rejected due to poor atmospheric transmission.  For the
17--24 \mum spectrum, data were rejected where the
transmission was less than 30\% of the maximum (atmosphere
and filter) based on observations of HR 2990.  Additionally,
data between 10.1 and 11.8 \mum have been combined into a
single bin to improve the signal-to-noise ratio.

Each spectrum was flux calibrated by dividing it by the
spectrum of HR 2990 with N = $-$1.24 and Q = $-$1.21
(Tokunaga~\markcite{sil-to84}1984) and by assuming a
blackbody temperature of 5000 K.  For the 8--13 \mum
spectrum, observations of \ir08 were all made within 0.1
airmasses of the observation of the standard.  For the 17--24
\mum spectrum two groups of observations of \ir08 were
calibrated with two separate observations of the standard,
such that the same condition on the difference in airmass was
obtained, and they were subsequently combined.  No further
corrections for airmass were applied. Both the object and
standard spectra were calibrated in wavelength prior to
division by the standard by means of higher spectral order
measurements of emission lines emitted in the range 2.0--2.4
\mum by a Kr lamp as described by Hanner, Brooke, \& Tokunaga
~\markcite{sil-ha95}(1995).  We estimate the absolute flux
calibration to be better than 15\% for the 10 \mum spectrum
and better than 30\% for the 20 \mum spectrum, and the
wavelength calibration to be better than half a resolution
element.

\subsection{Photometry}

In order to extend wavelength coverage to shorter
wavelengths, IRAS 08572+3915 was also observed on the night
of UT 1997 January 6 under good conditions using NSFCAM on
the NASA 3 m Infrared Telescope Facility (IRTF) on Mauna Kea,
Hawaii.  The camera detector was a 256 $\times$ 256 pixel
InSb array and the camera plate scale was set to 0\farcs15.
Three filters were used with central wavelengths at (and
spectral FWHMs of) 2.21 (0.39), 3.78 (0.59), and 4.77 (0.23)
\micron.  Ten pointings of 10 to 20 s each were made in each
filter with a 10\arcsec~ N-S offsets between pairs of images,
and with each pair offset from the other four using a domino
5 pattern of offsets with a 5$\sqrt{2}$\arcsec~ diagonal
spacing.  The standard star HD 22686 was also observed with
five pointings for each filter in a domino 5 pattern and
within 0.2 airmasses of the observations of IRAS 08572+3915.
Bias subtraction was performed by subtracting alternate
pointings for the galaxy, and subtracting the average of the
four alternative frames for the standard.  The difference
images were flatten using the median normalized difference of
a 1 s and a 0.5 s dome flat for the 2.21 \mum data, and by
using the median of the normalized images for the 3.78 and
4.77 \mum data.  Hot pixels were replaced by the median of
their nominally performing neighbors in a 3 $\times$ 3 pixel
box.  The difference images were shifted according to their
centroids and summed.  Flux densities for the galaxy in each
filter were estimated by comparing the total number of counts
within a 4\farcs5 diameter synthetic circular aperture above
the average counts in a concentric annulus of inner diameter
4\farcs5 and outer diameter 5\farcs4 to the same measure of
the standard star data scaled to match in integration time.
The 2.21 \mum magnitude of the standard was taken to be
7.195 (Elias et al.~\markcite{sil-eli82}1982) and assumed to
be 7.2 for the observations at the other two wavelengths.

\section{Results}

\begin{figure}
\plotfiddle{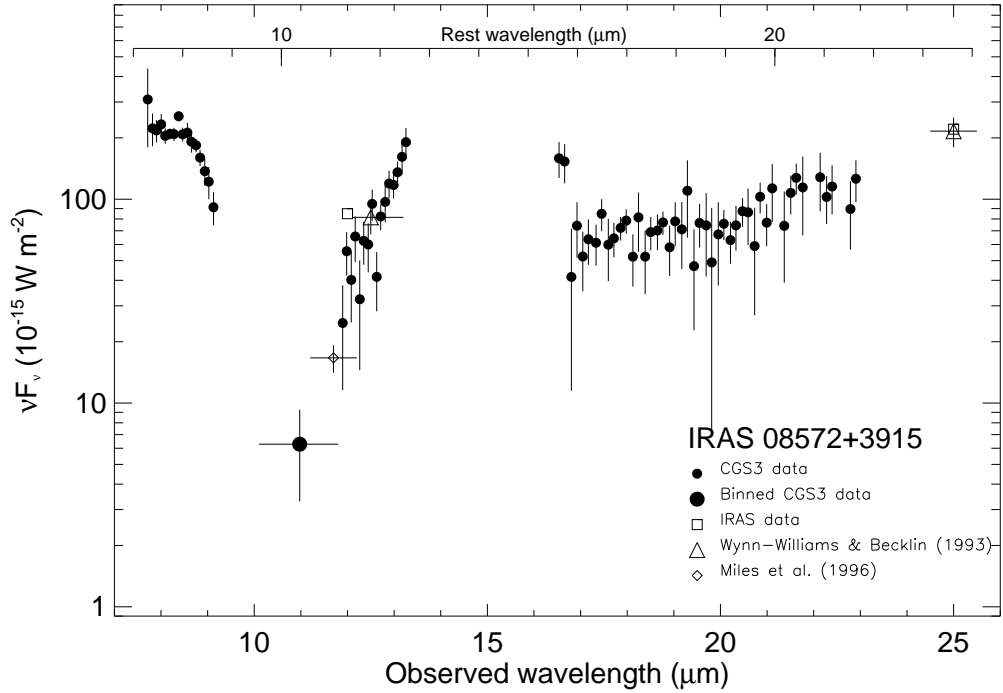}{350pt}{0}{120}{120}{-231}{0}
\caption[figure1.eps]{ The 10 and 20 \mum CGS3 spectra of
IRAS 08572+3915 are presented along with photometry from the
literature. The observed wavelength is given on the upper and
lower bounding box scale, and the rest wavelength
$(1+z=1.058)$, based on the millimeter spectroscopy given in
Sanders et al.~\protect\markcite{sil-sa89}(1989), is shown
near the top of the figure.  The small circles are independent
spectral data points, while the large circle is data binned
between 10.1 and 11.8 \mum (observed wavelength).  The
photometric data are from Wynn-Williams \&
Becklin~\protect\markcite{sil-wy93}(1993) (open triangles) and
Miles et al.~\protect\markcite{sil-mil96}(1996) (open diamond)
in 5\farcs5 and 4\farcs6 apertures, respectively, and IRAS
data (open squares) from Soifer et
al.~\protect\markcite{sil-so89}(1989).  Vertical lines on data
points represent 1 $\sigma$ errors, while horizontal lines
indicate photometric passbands.}
\end{figure}

In Figure 1, our 8--13 and 17--24 \mum CGS3 spectra of \ir08
are presented along with  photometric
measurements at a variety of wavelengths.  Given the very
different spectral resolution and different passband coverage,
we find good agreement between the photometric measurements and
our spectra.  The agreement at 12.5 \mum allows a limit on
source variability to be set at less than 30\% over 5 years.

The data show strong silicate absorption centered near 9.7
\mum (rest wavelength), but there is no clear indication of
silicate absorption centered near 18 \mum (rest wavelength)
such as that seen toward the Galactic Center (McCarthy et
al.~\markcite{sil-mc80}1980).  No detectable evidence of the
11.3 \mum PAH emission feature (F(11.3\micron)$\leq$0.7$\times
10^{-15}$ W m$^{-2}$ 3$\sigma$) is observed.  Two other
spectral features commonly seen in starburst galaxies, the 8.6
\mum PAH feature and the 12.8 \mum [\ion{Ne}{2}] feature, have
not been measured due to the redshift of the source imposing a
coincidence with telluric O$_3$ and CO$_2$ bands,
respectively.

\begin{deluxetable}{lcccc}
\tablewidth{0pt}
\tablecaption{Near-Infrared Photometry for IRAS 08572+3915}
\tablecolumns{2}
\tablehead{\colhead{$\lambda$} & \colhead{$f_\nu$} \\
\colhead{(\micron)} & \colhead{(mJy)}}
\startdata
2.21 & 4.4$\pm$0.9 \nl
3.78 & 50$\pm$10 \nl
4.77 & 100$\pm$20 \nl
\enddata
\end{deluxetable}

Table 1 presents our new photometric data for IRAS 08572+3915
at 2.21, 3.78, and 4.77 \micron.  We estimate these data to
be accurate to 20\%.  The 3.78 and 4.77 \mum data confirm the
trend seen at shorter wavelengths that IRAS 08572+3915
displays a strong nonstellar continuum component that is
dominant at wavelengths longer than 2 \micron.

The measured flux densities reported in Table 1 are somewhat
higher than, although not incompatible with, those that have been
previously reported.  Observations using 5\arcsec~ apertures at
2.2 \mum have been reported by Sanders et
al.~\markcite{sil-sa88}(1988) and Young et
al.~\markcite{sil-yo96}(1996) who give 3.75 and 3.42 $\pm$ 0.07
mJy, respectively, for the NW source.  Carico et
al.~\markcite{sil-car90}(1990) and Zhou, Sanders, \&
Wynn-Williams~\markcite{sil-zh93}(1993) report 3.2 $\pm$ 0.2 and
3.0 $\pm$ 0.6 mJy respectively using 2\farcs5 synthetic apertures.
Using the SE source to perform differential photometry, we have
compared our 2.21 \mum image with the 2.15 \mum image of Sanders
et al.~\markcite{sil-sa97}(1997), and we find our new observations
to be $\sim$25\% brighter than their slightly shorter wavelength
observations; however this may be a result of the large H--K color
of this object (Carico et al.~\markcite{sil-car90}1990).  Our 3.78
\mum flux density is also 20\% higher than that reported by
Sanders et al.~\markcite{sil-sa88}(1988) (40 mJy).  We interpret
these results as tentative, but not conclusive, evidence for
variability in IRAS 08572+3915.

\section{Discussion}

\subsection{Infrared Galaxies with Deep Silicate Absorption}

IRAS 08572+3915 is the third infrared galaxy in which a deep
silicate absorption feature has been found, the others being
NGC 4418 (Roche et al.~\markcite{sil-ro86}1986) and Arp 220
(Smith et al.~\markcite{sil-sm89}1989).  These three sources
have other features in common in addition to the similarity of
their mid-infrared spectra.  All show relatively warm
far-infrared color temperatures and compact radio nuclei.
Additionally, they have \hbox{\lir} to \hbox{\lco} ratios that
are high compared to both infrared-emitting disk galaxies and
luminous infrared galaxies taken as a class.

The three galaxies are morphologically diverse.  NGC 4418 is a
normal barred spiral galaxy (SAB) (de Vaucouleurs et
al.~\markcite{sil-de91}1991), and \ir08 has a disturbed
morphology with a projected separation between optical nuclei
of $\sim$7 kpc (Sanders et al.~\markcite{sil-sa88}1988).  Arp
220 appears to be an advanced galaxy-galaxy merger with an
average 2.2 \mum surface brightness profile similar to an
elliptical galaxy (Wright et al.~\markcite{sil-wr90}1990), a
double nucleus with a projected separation of $\sim$360 pc at
2.2 \mum (Graham et al.~\markcite{sil-gr90}1990), and a much
obscured central region displaying a number of closely spaced
hot spots at optical wavelengths (Shaya et
al.~\markcite{sil-sh94}1994).

Spectra of the three galaxies at wavelengths shorter than 8
\mum exhibit a variety of phenomena.  At visible wavelengths
\ir08 and Arp 220 display the type of spectra characteristic
of a low-ionization nuclear emission region (LINER) (Veilleux et
al.~\markcite{sil-ve95}1995), while NGC 4418 is classified as
AGN-like by Armus, Heckman, \& Miley~\markcite{sil-ar89}(1989),
but with extremely weak emission features (Roche et
al.~\markcite{sil-ro86}1986).  At near-infrared wavelengths
Arp 220 shows a deep 2.3 \mum CO absorption (Goldader et
al.~\markcite{sil-go95}1995).  NGC 4418 shows strong $Q$-band
H$_2$ emission (Ridgway, Wynn-Williams, \&
Becklin~\markcite{sil-ri94}1994).  \ir08 is dominated by a red
continuum at 2.2 \mum (Goldader et
al.~\markcite{sil-go95}1995) and an unusually strong 3.4 \mum
dust absorption feature (Wright et
al.~\markcite{sil-wr97}1997), the first such observed in an
external galaxy.

Spectral energy distributions for the three galaxies are
plotted in Figure 2.  In the case of NGC 4418 we have excluded
from our plot data that appear as upper limits in the range
9.4--11.1 \micron.  The spectrophotometric data for Arp 220
are corrected for contamination by a starburst component
responsible for $\sim$2\% of the 8--1000 \mum infrared
emission by subtraction of the spectrum of NGC 7714 (Phillips,
Aitken, \& Roche~\markcite{sil-ph84}1984) scaled by a factor
of 0.3, following Smith et
al.~\markcite{sil-sm89}(1989). However, we have regridded the
spectrum of NGC 7714 to the sampling of the spectrum of Arp
220 rather than employing a smoothed version. We note that the
subtraction of the starburst component is important only near
the 9.7 \mum minimum and near 11.3 \micron; it leaves the rest
of the spectrum largely unchanged.

\subsection{Silicate Fitting}

As a first step toward interpreting the silicate
absorption features in the three galaxies, we attempted
to fit their 8--13 \mum spectra by a model consisting of
a power-law energy source (S$_\nu \propto \nu^{-\alpha}$)
behind a screen of cold dust.  The extinction curve we
used for the dust model, $\tau(\lambda)$, is that of
Mathis~\markcite{sil-ma90}(1990) for R$_V \sim 3$.  It
includes 9.7 and 18 \mum silicate features $\tau_{\rm
Sil}(\lambda)$, the shapes and relative strengths of
which are taken from Draine \& Lee (1984), plus a
component of continuous absorption, $\tau_{\rm
Cont}(\lambda)$, that is proportional to $\lambda^{-1}$
over the wavelength range of interest.  Mathis lists
values for his extinction curve at only a few
wavelengths; we generated a much more closely sampled
version of Mathis' extinction curve between 8 and 13 \mum
from the $\mu$ Cep emissivity curve given by Roche \&
Aitken~\markcite{sil-ro84}(1984).  For the 18 \mum
feature we have adopted a feature profile derived from
the new {\sl Infrared Space Observatory (ISO)}
observations of the absorption seen against circumstellar
carbon dust emission heated by Wolf-Rayet stars. Details
are given in the Appendix.  According to this extinction
model, the silicate feature component contributes 3 times
as much extinction at 9.7 \mum as the continuum
extinction but nothing outside the range 8--25 \mum
(i.e., $\tau_{\rm Sil}(9.7) = 3\times\tau_{\rm
Cont}(9.7)$).

Including both continuum and feature extinction allows us to
estimate the total extinction ($\tau_{\rm
Tot}(\lambda)=\tau_{\rm Cont}(\lambda)+\tau_{\rm
Sil}(\lambda)$) somewhat akin to the strategy of
Willner~\markcite{sil-wi77}(1977), as well as $\tau_{\rm
Sil}$(9.7), which is defined observationally (Aitken \&
Jones~\markcite{sil-ai73}1973) as $$\tau_{\rm Sil}(9.7) =
\ln\left(\frac{ F_\lambda(8)+F_\lambda(13)}{2\times
F_\lambda(9.7)}\right).$$ However, in the case of the three
galaxies in question, saturation of the feature causes
$F_\lambda(9.7)$ to be poorly determined, and it was
necessary to fit a silicate profile over the whole range
(7.7--13.2 \mum) to make any estimate at all.  Additionally,
as can be seen in Figure 1, the rest wavelength 13 \mum is
not observed for \ir08 due to its redshift; our estimate of
$\tau_{\rm Sil}(9.7)$ relies therefore upon an extrapolation
of the fit to 13 \micron.  We did not attempt to fit
simultaneously the 17--24 \mum data for the three galaxies.

In fitting our extinction model, we have employed a modified
version of Bevington's~\markcite{sil-be69}(1969) nonlinear
least-squares fitting routine provided with the data language
IDL version 4.0.  Three free parameters in combination with
our interstellar extinction model have been employed.  The
free parameters are the power-law exponent ($\alpha$), the
scaling, and the amount of extinction ($\tau_{\rm
Tot}(\lambda)$).  The fit has been made to data in 5 to 6 bins
chosen to have roughly equivalent signal-to-noise ratios.  On
each iteration of the fit, the model was binned in the same
manner as the data before the $\chi^2$ estimation was made.
In the case of \ir08, data between 9.2 and 10.1 \mum have been
excluded from the fit where telluric O$_3$ absorption makes
data less reliable.  In the case of NGC 4418, data appearing
as upper limits close to the 9.7 \mum minimum were not
included in the fit, leading to a somewhat lower estimate of
the silicate optical depth at 9.7 \mum than that given by
Roche et al.~\markcite{sil-ro86}(1986), while in the case of
Arp 220, we measure a somewhat deeper silicate feature than
that reported by Smith et al.~\markcite{sil-sm89}(1989).  We
ascribe the difference to our slightly different method of
subtracting the starburst component as noted in \S ~4.1.  For
both NGC 4418 and Arp 220 our estimates of $\tau_{\rm
Sil}(9.7)$ agree with the previous estimates to within 25\%.

The optical depths and power-law exponents from our best fits are
indicated in Figure 2, where $\tau_{\rm Sil}(9.7)$, $\tau_{\rm
Tot}(9.7)$, and $\alpha$ are as defined above.

\begin{figure}
\plotfiddle{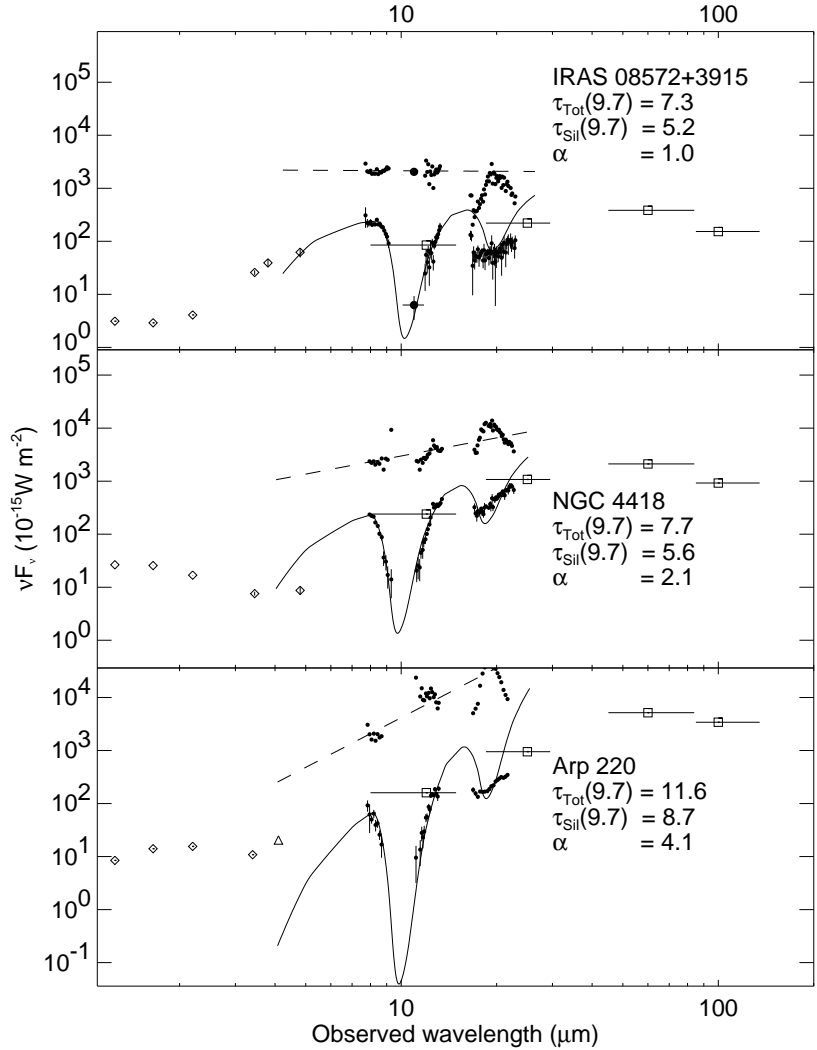}{372pt}{0}{100}{100}{-125}{30}
\caption[figure2.eps]{The 1.25--120 \mum spectral
energy distributions of the three galaxies with deep
silicate absorption features are shown.  The 8--24 \mum
spectroscopy (filled circles with error bars) is from
this work for IRAS 08572+3915, from Roche et
al.~\protect\markcite{sil-ro86}(1986) for NGC 4418, and
from Smith et al.~\protect\markcite{sil-sm89}(1989) for
Arp 220.  The 8--13 \mum spectral data for Arp 220 have
been corrected for a starburst component.  For each
galaxy, two series of points are plotted for the 8--24
\mum data.  The lower points are the observed data, while
the upper points (without error bars) represent the
result of deextinguishing the data to provide the
best-fit power-law spectral energy distribution over the
range 8--13 \mum (see \S ~4.2).  The dashed line is the
resulting best-fit power-law model without the effects of
extinction, and the solid line is the best-fit model
including the effects of extinction.  The broad-band
photometric points (open squares) between 12 and 100 \mum
are from Soifer et al.~\protect\markcite{sil-so89}(1989).
The 1.25--3.4 \mum photometry (open diamonds) for \ir08
and Arp 220 is taken from Zhou, Wynn-Williams, \&
Sanders~\protect\markcite{sil-sa93}(1993), with the
addition of our new 3.8 and 4.8 \mum photometry for IRAS
08572+3915 and a 4.1 \mum continuum point for Arp 220
from DePoy, Becklin, \&
Geballe~\protect\markcite{sil-dep87}(1987) (open
triangle). The 1.25--4.8 \mum photometric data for NGC
4418 were obtained with a 5\arcsec~ aperture at UKIRT
using the near-infrared photometer UKT9 on UT 1991
December 6 (Dudley~\protect\markcite{sil-dud97}1997).}
\end{figure}

\subsection{Problems with the Cold Screen Model for Deep Silicate
Infrared Galaxies }

Problems arise in attempting to explain the presence of
the deep silicate feature as foreground extinction by cold dust.

\begin{figure}
\plotfiddle{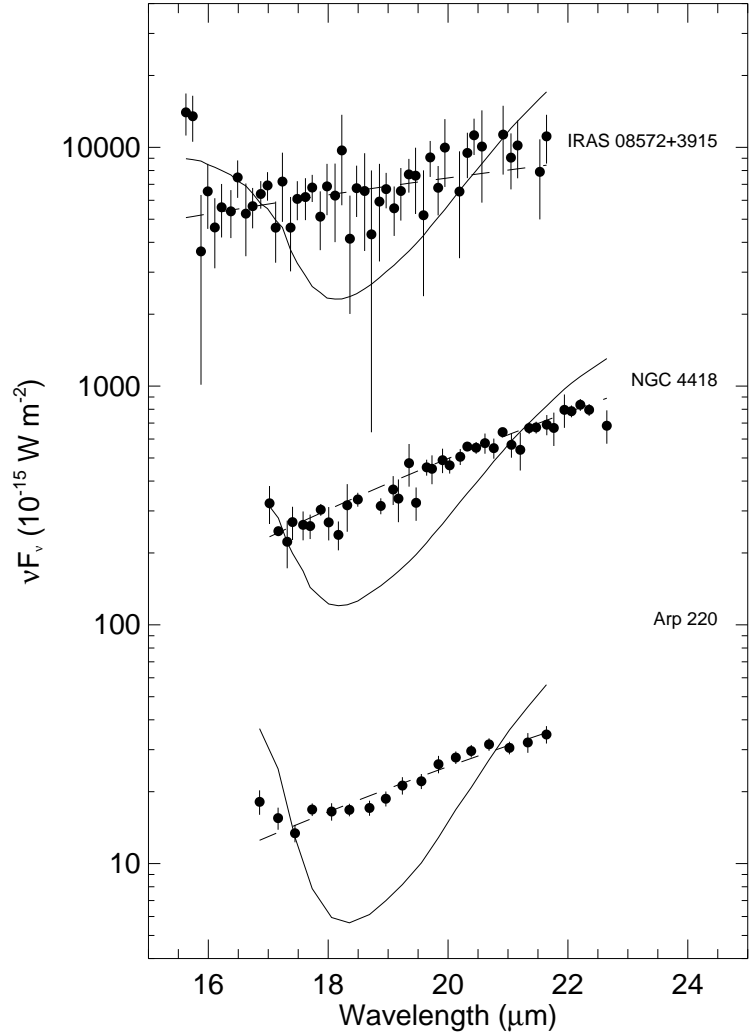}{350pt}{0}{120}{120}{-185}{0}
\caption[figure3.eps]{A detailed comparison of the 16--24
\mum spectra of each of the three galaxies is made with the
expected shape and depth of the 18 \mum silicate feature
(solid line).  The sources for the spectral data are the same
as in Figure 2.  In each case a power-law (dashed line) gives
a much better fit to the spectrum than that produced by
fitting a power-law extinguished by the amount of extinction
predicted from the 9.7 \mum feature.  For clarity, the
spectrum of IRAS 08572+3915 was shifted to its rest
wavelength, and scaled up by a factor of 100, while the Arp
220 spectrum was scaled down by a factor of 10.}
\end{figure}

First, there is no clear evidence for the predicted 18
\mum silicate absorption feature in any of the three
galaxies.  The difference is seen in Figure 2 and also in
more detail in Figure 3, which compares the observations
of the three galaxies with the expected shape and depths
of the 18 \mum silicate feature predicted from the
measured 9.7 \mum silicate optical depths.  Roche et
al.~\markcite{sil-ro86}(1986) have argued that in the
case of NGC 4418, the center of the 18 \mum feature may
be shifted to a shorter wavelength, possibly as a result
of differing dust composition, but given the larger
redshift of \ir08, such a shorter wavelength minimum
should be clearly evident in the 17--24 \mum spectrum of
\ir08 even if centered at 17 \micron.  Note that the
16--24 \mum spectrum of Arp 220, unlike that at 8--13
\micron, does not have a starburst component subtracted.
The weakness of the 18 \mum silicate feature in this
galaxy could be due partly to its being filled in by some
starburst emission.

Second, invoking a cold foreground screen of dust to explain
the appearance of the 8--13 \mum spectra of deep silicate
infrared galaxies runs into serious difficulty if the
background source is an extended starburst, as in most
high-luminosity infrared galaxies. The typical diameter of
starbursts in high-luminosity infrared galaxies is of the
order of a few hundred pc (Condon et
al.~\markcite{sil-co91}1991; Wynn-Williams \&
Becklin~\markcite{sil-wy93}1993).  To produce a silicate
optical depth of 5 in such a galaxy spectrum would require a
``wall" of at least this diameter with visual extinction of
$A_V \sim$ 75--90 mag (Aitken~\markcite{sil-ai81}1981) and a
column density of $N_H \sim 1.5\times 10^{23}$ cm$^{-2}$ if
the gas to dust ratio is standard.  This high value of the
column density must be maintained throughout the ``wall" or
the silicate feature will be filled in by emission from hot
dust in \ion{H}{2} regions and photo-dissociation regions
emerging through gaps in the ``wall.''  We would expect that
such an extended high column density screen should quickly
become clumped, leaving holes of lower column density.  We
thus believe that the presence of a deep silicate absorption
feature is incompatible with an extended starburst origin for
the power source in these galaxies, irrespective of the
presence or absence of PAH features in their spectra.

Third, the spectral energy distributions shown in Figure 2
are incompatible with the cold absorber model if spherical
symmetry is assumed for the sources.  For a hot object that
is completely surrounded by a cold dust shell, energy
absorbed in the UV through mid-infrared must be reradiated at
far-infrared wavelengths.  In a log $\nu$F$_\nu$ versus log
$\lambda$ plot, such as Figure 2, the reradiation process can
be visualized as a rightward horizontal shift along the log
$\lambda$ axis assuming that the spectral energy
distributions of the of the heating source and reradiating
source have approximately the same width in log $\lambda$.
But, as seen in Figure 2, the calculated source flux at 10
\mum matches or exceeds the maximum observed (i.e.,
reradiated) flux at far-infrared wavelengths in all three
sources.  In the case of \ir08 the excess is a factor of
5--6.  If we are to believe that \ir08 contains an object
with a power-law energy distribution given by the dashed line
in Figure 2, then four-fifths of its power must be lost to
space without being intercepted by dust.  Given the large
column density of dust in the line of sight to the core of
\ir08, we consider such a geometry unlikely.  Thus, this
effect also calls into question the appropriateness of
interpreting the deep silicate features as arising from pure
extinction.

None of the above problems by itself completely eliminates the
possibility that the 8--13 \mum spectra originate from cold
dust, but the doubt that they raise collectively encourages us
to look at other explanations for the spectra.  The most
promising alternative is discussed in the next section.

\subsection{The Protostar Analogue}

The only other known infrared astronomical objects that have
confirmed silicate absorption features as strong as those of
the three galaxies under discussion are certain luminous
``protostars" in galactic star-forming regions.  Prime
examples are W3 IRS5 (Aitken \& Jones~\markcite{sil-ai73}1973;
Willner~\markcite{sil-wi77}1977) and W33 A (Capps, Gillett, \&
Knacke~\markcite{sil-ca78}1978), which are thought to be new
stars (protostars rather than young stellar objects, although
at high mass this distinction breaks down due to rapid
evolution toward the main sequence) that are still deeply
embedded in the material out of which they have formed.  It
would be unreasonable to invoke a multitude of individual
protostars to explain the present data (because other phases
of star formation would dominate the mid-infrared spectrum in
a composite spectrum), but we can consider the physical
conditions in galactic protostars as an analogue for
understanding deep silicate infrared galaxies.

In such luminous protostars the deep silicate feature is
understood to arise not from pure absorption by cold dust, but
from a centrally heated optically thick dust envelope that has
a negative radial temperature gradient away from the power
source (Scoville \& Kwan~\markcite{sil-sc76}1976; Kwan \&
Scoville~\markcite{sil-kw76}1976).  We observe a deep silicate
feature because we are observing only to an optical depth of
order unity at any given wavelength and we see to deeper (and
therefore hotter) layers at 8 and 13 \mum than at 9.7 \micron,
where the dust opacity is larger.  The depth of the 9.7 \mum
silicate feature is thus a result of differences in the source
function at different optical depths of the dust envelope.
These differences can be especially large if the emission at
10 \mum is coincident with the Wien side of the Planck
function.

The problems with the cold screen model discussed in \S ~4.3 are
accounted for naturally in the Kwan and Scoville model.  The
weakness of the 18 \mum silicate absorption feature can be
understood because in the 20 \mum region, we are both looking
deeper into the dust envelope and looking at a wavelength range
that is closer to the peak of the Planck function, so that the
depth of the 18 \mum feature is less than would be expected for
pure extinction based on the 9.7 \mum depth. Further, only a
compact source is required so that covering an extended source
is no longer a difficulty.  Finally, energy is conserved in the
Kwan and Scoville model, so that it is no longer necessary to
postulate the loss of energy to space, particularly in the case
of \ir08.

We have examined the analogy with galactic protostar models
further by comparing the three galaxies' spectra with the grid
of spherical protostar models calculated by
Rowan-Robinson~\markcite{sil-row82}(1982). We have chosen these
calculations over later work (Efstathiou \&
Rowan-Robinson~\markcite{sil-ef90}1990; Adams \&
Shu~\markcite{sil-ad86}1986) because they rely upon fewer
geometrical assumptions but include the effect of scattering
with an assumed isotropic phase function, which can make an
important contribution to the emission near 2 \micron.

These calculations describe the transfer of radiation originating
from a central UV thermal source through a spherically symmetric
dust shell.  The inner radius of the dust shell is set by the
assumed dust sublimation temperature (1000 K), while the outer
radius is assumed to be a multiple of the inner radius (typically
1000).  The two most important parameters affecting the spectral
shape of the emergent radiation examined in these models are the
total optical depth at UV wavelengths and the change of density
with radius; some effect is also seen as a result of varying the
extent of the outer radius.  Because of the high optical depth at
UV wavelengths in these models, they are insensitive to the
precise spectral energy distribution of the central energy source,
and at radii greater than the dust sublimation radius, essentially
all the radiation is dust emission.  In the wavelength region
5--24 \mum the model dust opacity used by
Rowan-Robinson~\markcite{sil-row82}(1982) is in essence the same
as that employed in this paper in that the ratios of silicate to
continuum opacity at the centers of the 9.7 \mum silicate and the
18 \mum silicate are nearly identical and the wavelength
dependence of the continuum opacity has the same functional form.
There is a difference in that the spectral shape of the 9.7 \mum
silicate feature in the Rowan-Robinson models is assumed to be
Gaussian rather than asymmetric in wavelength. It is also broader
in wavelength so that the relevant comparison is not with the
detailed spectral shapes of the silicate features, but rather with
their depths.

\begin{figure}
\plotfiddle{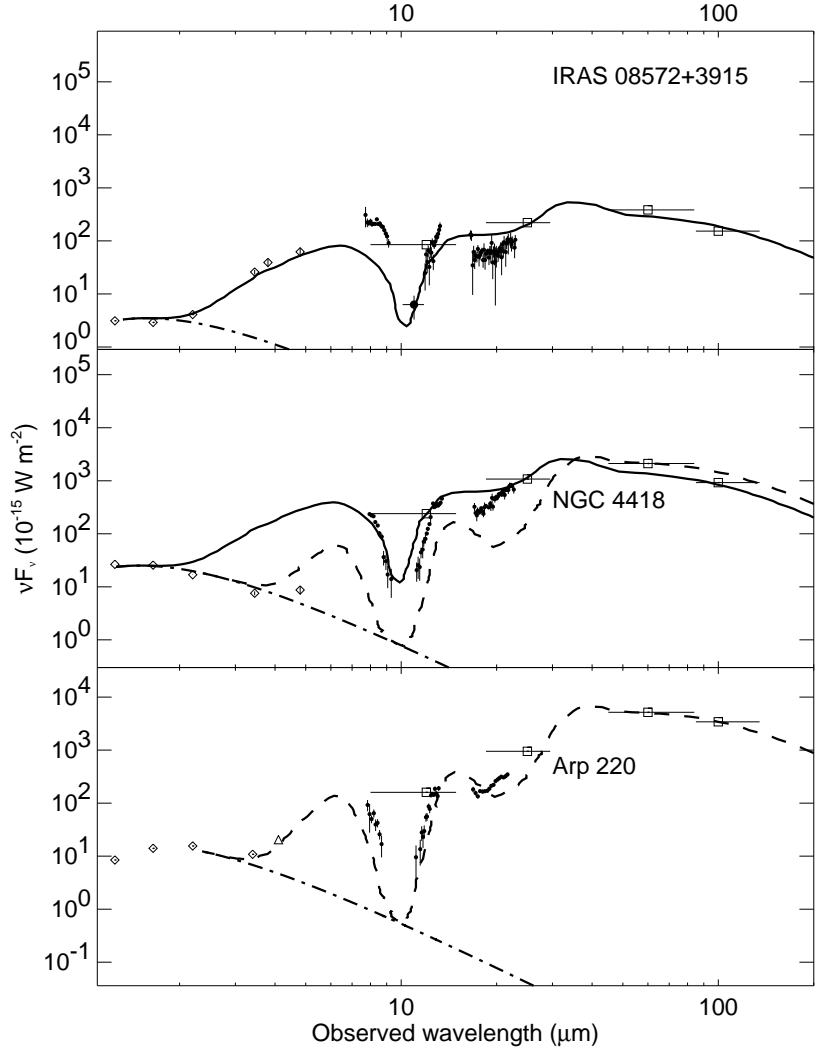}{372pt}{0}{100}{100}{-125}{30}
\caption[figure4.eps]{Observed spectral energy
distributions of \hbox{IRAS 08572+3915,} NGC 4418, and Arp 220
are compared with scaled-up protostar models. The solid curves
correspond to a calculation with $\tau_{UV}$ = 200, and the
dashed curves correspond to a calculation with $\tau_{UV}$ =
500. The dot-dashed curves are 2500 K grey bodies that have
been included to represent the emission from the stars in the
galaxies that fall within the photometric beam.}
\end{figure}

In Figure 4 we compare the 1.25--100 \mum spectral energy
distributions of the deep silicate infrared galaxies, as
presented in Figure 2, with two of Rowan-Robinson's models that
differ in only one property, the UV optical depth
($\tau_{UV}=\tau(\lambda<0.1$\micron)).  It is clear that the
overall match to the spectral energy distribution of the deep
silicate infrared galaxies is quite good.  We find the closest
agreement between model and observation for Arp 220, indeed both
the 9.7 and apparent 18 \mum feature depths are well matched.
NGC 4418 appears bracketed by the two models, while \ir08 shows
somewhat stronger emission at 8 \mum than the $\tau_{UV}=200$
model would predict.  The relative weakness of the 18 \mum
silicate feature compared to the 9.7 \mum silicate feature in
the $\tau_{UV}$ = 200 model is in particular agreement with the
\ir08 observations, while the agreement from 2 to 5 \mum is
quite encouraging.

The density profiles of both models (i.e., with $\tau_{UV}$ =
200 and $\tau_{UV}$ = 500) are proportional to $r^{-1}$.  We
note that models with density profiles proportional to $r^{-2}$
do not match the depths of the silicate features nor the overall
spectral energy distributions of deep silicate infrared
galaxies up to the maximum $\tau_{UV}$ calculated
($\tau_{UV}$ =500), while models with uniform density tend to
overestimate the strength of the 18 \mum silicate feature.

\subsection{Physical Scales}

The models introduced in the \S ~4.4 were originally
calculated for galactic protostars with $L_{\rm IR}$ $\sim$ 10$^5$
\lsun, but they may be scaled (Zeljko \&
Elitzur~\markcite{sil-ze97}1997) to the luminosity range
10$^{11}$--10$^{12}$ \hbox{\lsun} by maintaining a constant
emergent surface brightness and column density while the
radius is increased.  Thus the average density reduces as
1/$R$, while the mass and luminosity increase as $R^2$, where
$R$ is a characteristic size.

Given that the detailed agreement between the models and the
observations is only approximate, the following derived
parameters should be taken as indicative only.  The $\tau_{UV}$
= 200 model has an emergent surface brightness of $1.4 \times
10^{-3}$ W m$^{-2}$ sr$^{-1}$ at 25 \mum at the outer radius of the
model.  \ir08 produces $2.2\times 10^{-13}$ W m$^{-2}$ at 25 \mum
at a distance of 230 Mpc.  Thus by scaling the model to match
at 25 \micron, an angular source size of $1.6\times 10^{-10}$
sr is deduced; the angular outer radius is 1\farcs 5, or 1800
pc.  The inner dust sublimation radius is 1.8 pc.  For an UV
optical depth of 200, $A_V$=52 so that for $N_H = 18\times
10^{21}A_V$ cm$^{-2}$ upon integration, the total gas mass for
the model is $\sim$10$^{10} M_\odot$.

\begin{deluxetable}{lcccc}
\tablewidth{0pt}
\tablecaption{Scaled Protostar Models}
\tablecolumns{5}
\tablehead{ & \colhead{\ir08} & \multicolumn{2}{c}{
NGC 4418} & \colhead{Arp 220} \\
Property & \colhead{$\tau_{UV}$=200} & \colhead{$\tau_{UV}$=200} &
\colhead{$\tau_{UV}$=500} & \colhead{$\tau_{UV}$=500} }
\startdata
$d$ (Mpc) & 230 & \multicolumn{2}{c}{ 27 } & 73 \nl
$L$ (\lsun) & 1.3$\times 10^{12}$ & \multicolumn{2}{c}{
1.3$\times 10^{11}$} & 1.5$\times 10^{12}$ \nl
$\lambda$ (\micron)& 25 & 25 & 60 & 60 \nl
$r_{\rm inner}$ (pc) & 1.7 & 0.44 & 0.46 & 2.1 \nl
$r_{\rm inner}$ (mas) & 1.5 & 3.3 & 3.5 & 5.8 \nl
$r_{\rm outer}$ (arcsec) & 1.5 & 3.3 & 3.5 & 5.8 \nl
$r_{\rm radio}$ (mas) & 40 & \multicolumn{2}{c}{ 200} & 120\tablenotemark{a} \nl
$M_{\rm g}$ (10$^9$ \msun) & 10 & 1 & 2.5 & 25 \nl
$M_{\rm g}$(CO) (10$^9$ \msun) & 7 & \multicolumn{2}{c}{ 1.4} & 20 \nl
$2\sqrt{2\ln{2}} \times \sqrt{\frac{GM}{6R}}$ (\kms)\tablenotemark{b} & 150 & 95 & 150 & 220 \nl
FWHM(CO) (\kms) & 190 & \multicolumn{2}{c}{ 120} & 360 \nl
References & 1, 3 & \multicolumn{2}{c}{ 2, 4} & 1, 4 \nl
\enddata
\tablenotetext{a}{The radio size for Arp 220 is for the western
component only.}
\tablenotetext{b}{The virial
velocity variance will be $(GM)/(2R)$ for a system with mass
$M$ and radius $R$  (Mihalas~\protect\markcite{sil-mi67}1967
equation 14-14).  This must be divided by a factor of 3 to give
the line-of-sight variance.}
\tablerefs{Radio sizes at (1) 3.6 cm from Condon et al.~\protect\markcite{sil-co91}(1991) and (2) 20 cm Condon et al.~\protect\markcite{sil-co90}(1990); CO data from (3) Sanders
et al.~\protect\markcite{sil-sa89}(1989) and (4) Sanders, Scoville, \& Soifer~\protect\markcite{sil-sa91}(1991).}
\end{deluxetable}

\begin{deluxetable}{lccc}
\tablewidth{0pt}
\tablecaption{Planck Temperatures and Sizes at 8 and 13 \micron}
\tablehead{Property & \colhead{\ir08} & \colhead{NGC 4418} & \colhead{Arp 220} }
\startdata
$T_{8-13\micron}$ (K) & 370 & 280 & 200 \\
$R_{8-13\micron}$ (pc) & 2 & 0.6 & 3 \\
\enddata
\end{deluxetable}

In Table 2 we compare physical sizes, gas masses, and
expected gas velocities derived from scaling the
Rowan-Robinson~\markcite{sil-row82}(1982) models as shown in
Figure 4 with those deduced from observations of 3.6 or 20 cm
continuum and millimeter CO(1$\rightarrow$0) emission.  For
reference, observed galaxy distances and infrared
luminosities are given in the first two rows, while row 3
identifies the wavelength at which the scaling of the models
was performed. Rows 4 and 5 give the inner radius (the dust
sublimation radius) where the dust temperature is 1000 K in
pc and milliarcseconds (mas), and row 6 gives the model outer
radius in arcseconds.  Row 7 gives the observed radio
size. Row 8 gives the gas mass based on the model, while row
9 gives the gas mass based on observations of
CO(1$\rightarrow$0). Row 10 gives the expected gas
line-of-sight velocity full width at half-maximum (FWHM) using
the model outer radius and the model gas mass under the
assumption of a virialized pressureless system with a
Gaussian line profile, and row 11 gives the observed CO
velocity FWHM.  Row 12 gives references for the radio and CO
data.  Given the approximate nature of the matches between
the spectral energy distributions of the galaxies and the
scaled-up protostar models, we find the comparison of model
and observation to be quite encouraging.  A quite similar
model applied to Arp 220 by Rowan-Robinson \&
Efstathiou\markcite{sil-row93}~(1993) leads to a very
similar estimate for the inner radius.  Their model has a
smaller outer radius and thus has a smaller gas mass.

In addition to the physical scales that can be estimated from
the model, there is another estimate of the
size of the warm dust emitting region that can be made directly
from the spectral data.  Under the assumption that we are
observing to an optical depth of order unity throughout
the 8--13 \mum spectral region, we will be observing down to
the same physical surface if the dust opacity is the same.  In
Mathis's~\markcite{sil-ma90}(1990) extinction curve, the opacities
at 8 and 13 \mum are very similar, so it is possible to estimate
the dust temperature at that layer.  Since the emission is from
an optically thick dust layer, the source size will be close to
that of a blackbody of that temperature.  In Table 3 we give
the color temperatures and radii of the blackbodies that
reproduce the 8 and 13 \mum flux at the distance of each
source for comparison with the values derived from the
models.  As noted in \S ~4.2, our estimate of the rest
wavelength 13 \mum emission for \ir08 is based upon an
extrapolation of our fit of the interstellar extinction curve
shown in Figure 2.  The results of Tables 2 and 3 are consistent
in that $R_{8-13\micron}$, where the dust temperature is
200--400 K, is in each case larger than $r_{\rm inner}$ where the
dust temperature is by definition 1000 K.

The most important aspect of this exercise is that it indicates
that the power that is emitted predominantly at 60 \mum in deep
silicate infrared galaxies conforming to this model is
originally produced inside of a region with $r_{\rm inner}$ in
the range 0.4--2 pc (Table 2).  It is this feature of the model
that most distinguishes it from an extended starburst model,
in which the emitting dust is optically thin (at infrared
wavelengths) and appears as a comparatively large solid angle
region of comparatively low surface brightness.

\subsection{The Nature of the Power Source}

In \S ~4.5 we argued that the spectral
energy distributions of the deep silicate infrared galaxies can
be fitted with a model where $10^{11}$--10$^{12}$ \lsun~ is
generated within a region only a few pc in size.  In this
section we discuss the nature of the power source.

Because of the high optical depths of dust in our suggested
model, we would not expect to see any spectroscopic features
associated with the sources that are produced by radiation that
is highly attenuated by dust.  Optical and UV lines that arise
inside the dust sublimation radius would be attenuated by
50--200 mag.  Since the source UV radiation is efficiently
converted to $T$=1000 K thermal emission at the dust
sublimation radius, PAH emission is not expected regardless of
the hardness of the source spectrum since there is little
short-wavelength radiation remaining to excite emission or
destroy the carriers at the radii we observe in the 8--13 \mum
region.  Further, we would not expect to observe variations on
a timescale shorter than the light-crossing time of about a
decade (see Table 3) irrespective of the intrinsic variability
of the original power source.

With this much power in a small volume we are almost certain
to be seeing the manifestations of an AGN of some kind,
although Terlevich et al.~\markcite{sil-te93}(1993) consider
that a starburst could be this compact.  Most AGN models
involve disklike accretion onto a black hole from a scale of
$\sim$3 $\times R_S$ and larger, where $R_S$ is the Schwarzschild
radius of the black hole while short-timescale variability of
known AGNs indicates sizes of order 0.05 pc.  Guided by
AGN models of this type, we might expect the inner regions of
the source to have a disklike structure.  We emphasize,
however, that the dust producing the 8--13 emission from deep
silicate infrared galaxies is unlikely to be very disklike
itself; if the infrared radiation were coming directly from a
hot disk seen partially face on, we would not see a deep
silicate feature. The compact disk models of Pier \&
Krolik~\markcite{sil-pi92}(1992) suffer from the difficulties
that (1) they present the sorts of spectral energy
distributions observed in deep silicate infrared galaxies for
a range of viewing angle much smaller than 1$^\circ$, and (2)
the optical spectrum should {\sl clearly} reveal the narrow
line region.  The first objection might be overcome by
noticing that thus far there are only three known deep
silicate infrared galaxies.  This argument could hold for NGC
4418 taken by itself as a member the class of luminous
infrared galaxies and assuming that all luminous infrared
galaxies are powered by AGNs. \ir08 and Arp 220 are, however,
an important fraction of the bright ultraluminous infrared
galaxies, and so for them compact disk models are an unlikely
explanation even if all ultraluminous infrared galaxies were
configured this way, without invoking the missing narrow line
region argument.  The tapered disk models of Efstathiou \&
Rowan-Robinson~\markcite{sil-ef95}(1995) do not suffer as
severely from restrictions on viewing angle
(Efstathiou~\markcite{sil-ef96}1996), but again the optical
spectra should reveal a {\sl powerful} narrow line region,
which is not the case for the known deep silicate infrared
galaxies.  More elaborate models such as embedded disks (along
the lines of the protostar models of Adams \&
Shu~\markcite{sil-ad86}(1986) are probably not required by the
present data, but neither are they ruled out.

The model we propose here lends some support to the proposal
of Sanders et al.~\markcite{sil-sa88}(1988) that ultraluminous
infrared galaxies are a part of an evolutionary sequence that
results in optically visible quasars.  If the AGNs that we
suggest as the power source for the deep silicate infrared
galaxies were to clear some of the enshrouding gas and dust,
or alternatively, if the enshrouding gas and dust were to
collapse to a disk on a timescale of $\sim$5 Myr, then a
quasar could result.  Mkn 231, which shows a Seyfert 1
optical spectrum and a silicate absorption feature (Roche,
Aitken, \& Whitmore~\markcite{sil-ro83}1983) that is weaker than
those discussed here, might be a transitional object.
Testing this possibility will require mid-infrared
spectroscopic observations of a substantially larger number
of ultraluminous infrared galaxies.

Sturm et al.~\markcite{sil-st97}(1996) have presented {\sl
ISO} observations of the strengths of the emission lines from
Arp 220 and present two lines of evidence to suggest that
this galaxy is powered by massive stars rather than by an
AGN.  We believe that their evidence does not conflict with
the model we have suggested in this paper.  Their first
argument is that they see no high excitation emission lines
such as would be produced by an AGN.  In our model, such high
excitation line emission is hidden within a very thick dust
shell and would not be expected to be readily observed.
Their second argument is that the line ratios of certain
hydrogen and sulfur lines indicate that they come from highly
reddened \ion{H}{2} regions; when the recombination lines are
dereddened, they imply Lyman continuum luminosities
consistent with all the galaxy's power arising from
\ion{H}{2} regions.  Our first comment is that their estimate
of the extinction is based partly on a comparison of
Br$\alpha$ and Br$\beta$ observed using a
14\arcsec$\times$20\arcsec~aperture with Br$\gamma$ observed
using a 3\arcsec$\times$15\arcsec~ aperture (Goldader et
al.~\markcite{sil-go95}1995).  If some of the recombination
lines come from an extended region, as is suggested by the
factor of 5 discrepancy between the {\sl ISO} measurements of
Br$\alpha$ and the UKIRT observations of Depoy et
al.~\markcite{sil-de87}(1987) made with a 10\arcsec~
aperture, then some of the evidence for high extinction may
be discounted.  The large difference between the remaining
estimates of $A_V$, namely, 30 mag for the Br$\alpha$ to
Br$\beta$ ratio and the limit of $\geq$59 mag from the limit
on the ratio of the [\ion{S}{3}] lines suggest that while the
light from massive stars implied by these estimates of the
Lyman continuum luminosity {\sl may} be sufficient to account
for the infrared luminosity of the whole galaxy, it is
equally possible that it contributes only a fraction of the
total luminosity.

While our model of an AGN deeply embedded in a spherical dust
shell provides an attractive explanation for both the depth
of the 9.7 \mum silicate feature and the overall spectral
energy distribution of the deep silicate infrared galaxies,
we concede that it cannot by itself explain all the nuclear
spectral features of these galaxies described in \S ~4.1.
Also, in all three galaxies there is evidence for radio
continuum emission on a scale larger than the dust
sublimation radius. Some of this extended radio emission
could arise from a nascent pair of jets that has not been
resolved in VLA observations.  It could also be due to
supernova remnants from a starburst that existed before the
AGN became dominant.  This would be consistent with the deep
2.3 \mum CO absorption, indicative of the presence of evolved
supergiant stars, observed by Goldader et
al.~\markcite{sil-go95}(1995) in the case of Arp 220.  In Arp
220, the presence of two compact sources in both the radio
(Norris~\markcite{sil-no88}1988; Condon et
al.~\markcite{sil-co91}1991) and at 12.5 \mum (Miles et
al.~\markcite{sil-mil96}1996) offers the attractive
possibility that the AGN and the starburst suggested by PAH
emission seen in the spectrum of Smith et
al.~\markcite{sil-sm89}(1989) may be physically distinct.
The detection of compact dust emission at 2.7 mm (Scoville et
al.~\markcite{sil-sc91}1991) and mas-scale radio structure
with very a high brightness temperature (Lonsdale et
al.~\markcite{sil-lo93}1993) at the western source suggest
that this may be the deep silicate source since, within the
context of our model, such indications of phenomena differing
from what is typical of spatially extended star formation
might be expected.  This prediction can be tested by
higher spatial resolution 8--13 \mum spectroscopic
observations.  The optical emission lines in the other
galaxies might be produced by an as-yet undetected second
nucleus, a {\sl small} leakage of radiation from the AGN's
dust shroud, or by the effects of more penetrating X-ray
radiation that are not included in our model.

\section{Conclusions}

We have presented new mid-infrared spectroscopy of the
ultraluminous infrared galaxy \ir08 and have found the 8--13
\mum spectrum to be dominated by a very strong silicate
absorption feature.  The 17--24 \mum spectrum does not show any
clear indication of the predicted 18 \mum silicate feature.
\ir08 is found to be a member of a group of deep 9.7 \mum
silicate infrared galaxies that include NGC 4418 and Arp 220.

We have presented fits of an arbitrary power law extinguished
by interstellar extinction for all three galaxies and
have shown that the calculated source flux in the 8--13 \mum
region is too high if cold shell absorption in the 8--13 \mum
region produces the far-infrared emission.  This difficulty, in
combination with the arguments that an extended cold screen is
unworkable and that there is little evidence for the 18 \mum
silicate feature in absorption in any of the sources, leads us
to try to fit a model along the lines of protostar models as a
simpler and qualitatively better explanation of the data.

We have used two protostar models that differ only in
optical depth to characterize the sources and make comparisons
with observation at other wavelengths.  We find that the
observed gas mass and velocity dispersions are fairly consistent
with model predictions.

Both the scaling of published protostar models and the
assumption that the optical depth at 8 and 13 \mum is of order
unity lead to sizes for the warm dust emission region that are
smaller than a few pc.

We have argued that if the sizes derived either from the
protostar models or by means of the blackbody assumption are
correct, then it is very unlikely that deep silicate infrared
galaxies are powered by star formation.

On the basis of these arguments, we propose tentatively that
for the three galaxies considered here, a substantial portion
of their infrared emission can be ascribed to AGNs nearly
completely surrounded by optically thick dust shells with hot
dust near the center and negative radial dust temperature
gradients.

\acknowledgments

We would like to thank the UKIRT staff (Joel Aycock,
Dolores Walther, and Thor Wold) for assistance at the
telescope, and Tom Geballe and Gillian Wright for
fruitful discussions and sharing spectral data in advance
of publication.  We also thank Bill Golisch, who ran the
IRTF during our observations.  This work has benefited
from discussion with and/or sharing of work in progress
by many members of the Institute for Astronomy.  A few
are Dave Sanders, Bob Joseph, Klaus Hoddap, Josh Barnes,
George Herbig, John Hibbard, Alan Tokunaga, Ken Chambers,
Jeff Goldader, Joe Jensen, and Jason Surace; we thank
these and others.  We thank John Dudley for a critical
reading of an early draft of this paper.  This work has
been partially supported by NSF grant ASTR-8919563 and
NASA grant NAGW-3938.  The NASA/IPAC Extragalactic
Database (NED) has been a constant aid.  It is run by the
Jet Propulsion Laboratory, California Institute of
Technology, under contract with NASA.  This work has also
made use of NASA's Astrophysics Data System Abstract
Service.

\appendix

\section{The Spectral Shape of the 18 \mum Silicate 
Feature}

The shape of the interstellar 18 \mum silicate feature is
poorly determined due to its relatively lower peak absorption
and greater breadth relative to the 9.7 \mum feature.  The
advent of {\sl ISO} has made it possible to determine the shape
of the profile with much greater precision than before.  In
this Appendix we derive the profile of the 18 \mum silicate
feature from {\sl ISO} Short Wavelength Spectrograph 3--30 \mum
spectra of Wolf-Rayet stars published by van der Hucht
et al.~\markcite{sil-va96}(1996).  The advantage of using these
stars' spectra as templates for the 18 \mum silicate absorption
feature is that the underlying energy distribution of the star
in the 3--30 \mum range arises from hot carbon dust and is
therefore smooth and free of any intrinsic silicate emission
features.  

To derive the shape of the 18 \mum feature, we have
modeled the carbon dust emission against which it is observed
as originating in shells of dust at a range of temperatures
following Williams, van der Hucht, \&
Th\'{e}~\markcite{sil-wil87}(1987) in a simplified manner.
The carbon dust emission from three Wolf-Rayet stars, WR 98a,
WR 112, and WR 118, was modeled as arising from dust with
emissivity proportional to $\lambda^{-1}$ in 10
shells with $T$ ranging from 350 to 1500 K in equal intervals.
WR 48a and WR 104, whose spectra were also reported by van der
Hucht et al.~\markcite{sil-va96}(1996), have not been modeled
in the first instance because our model is too simple to give
an adequate fit, and in the second instance, because the
extinction is too low to make a useful estimate of the shape
of the 18 \mum silicate feature.  The contribution of each
shell at 35 \mum was proportional to a power of the shell
temperature.  The interstellar extinction outside of the 18
\mum silicate feature toward the hot carbon dust was taken to
have the wavelength dependence employed throughout this paper,
namely, the curve given by Mathis~\markcite{ma90}(1990)
supplemented by the 8--13 $\mu$ Cep emissivity curve given by
Roche \& Aitken~\markcite{ro84}(1984).  The amount of
extinction for each of the stars, WR 98a, WR 112, and WR 118,
was estimated by fitting the 9.7 \mum silicate feature giving
$\tau_{\rm Sil}(9.7)$ of 0.68, 0.63, and 0.62, respectively.
Bevington's~\markcite{sil-be69}(1969) CURVFIT program was
employed to estimate the continuum flux density and the
exponent of the dust shell temperature required to give the
best fit to the observations in the wavelength ranges
4.5--5.8, 6.4--13.2, and 25.0--29.0 \micron.  Using these
parameters, the carbon dust continuum emission was estimated
over the whole range 3--30 \mum for each star.  We then used
this derived carbon dust emission to estimate the extinction
in the 18--25 \mum range.  The derived extinction curves of WR
98a and WR 112 were adjusted to match WR 118 at 9.7 \micron, and
the average extinction as a function of wavelength was
computed from these three estimates.

\begin{figure}
\plotfiddle{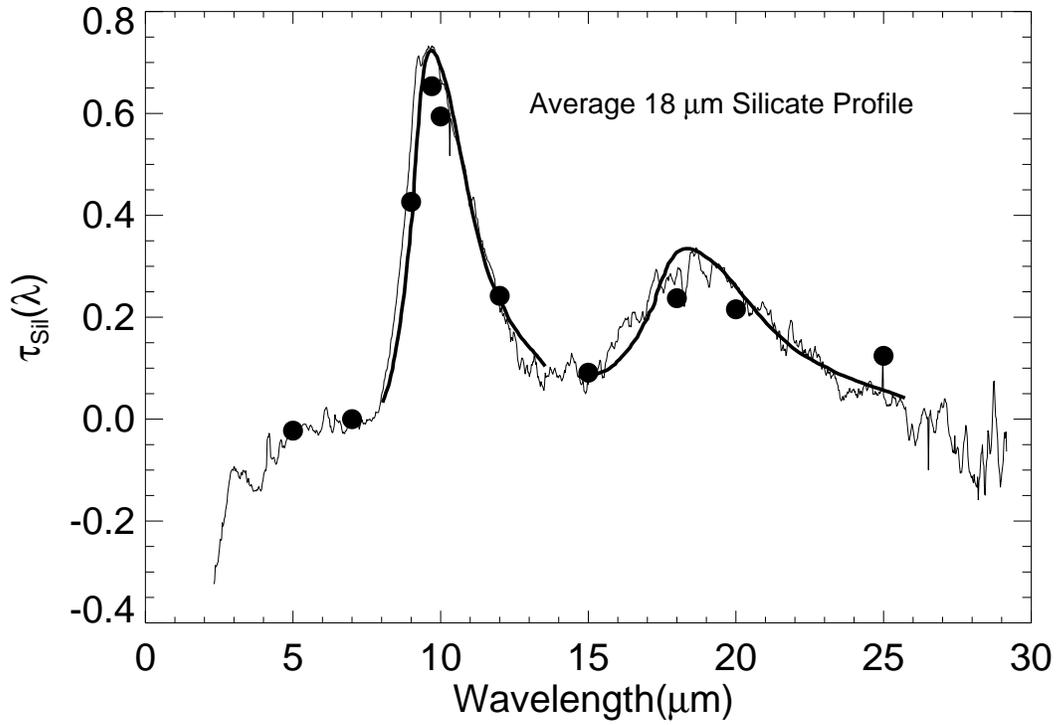}{350pt}{0}{120}{120}{-231}{0}
\caption[figure5.ps]{The average 18 \mum silicate feature
optical depth along the lines of sight toward three Wolf-Rayet
stars is presented.  The thin line gives our derived estimate
of the 18 silicate feature optical depth as a function of
wavelength.  Our adopted silicate feature profiles are shown as
thick lines.  The filled circles show the curve given by
Mathis~\protect\markcite{sil-ma90}(1990) after subtraction of
the continuous extinction.}
\end{figure}

Figure 5 presents the average results of our model fits.  The
vertical scale is silicate feature optical depth after removal
of the continuous extinction component, and the horizontal
scale is wavelength.  Our derived 18 \mum silicate profile is
given by the thin line in Figure 5.  The thick lines are the
smooth estimates of the silicate features that we adopt.  The
one peaking at 9.7 \mum is our adopted $\mu$ Cep emissivity
curve from Roche \& Aitken~\markcite{sil-ro84}(1984), and the
one centered at 18 \mum is our new estimate of the shape of
the 18 \mum silicate feature.

We have also plotted the extinction values tabulated by
Mathis~\markcite{sil-ma90}(1990) less the continuous
extinction and scaled to the appropriate optical depth.  The
differences seen between our adopted curve and Mathis's curve
(filled circles) near the peak of the 9.7 \mum feature are due
to Mathis's reliance on the quasi-empirical curve of Draine \&
Lee~\markcite{sil-dr84}(1984), which attempts to model the
silicate emission observed in the Trapezium region of the
Orion Nebula.  We find that the interstellar 18 \mum silicate
feature may also be more peaked than that employed by
Mathis~\markcite{sil-ma90}(1990).  We do not have the data to
estimate how much the ratios of the strengths of the silicate
features vary with respect to each other or how the feature
strengths vary with respect to the continuous extinction on
different lines of sight.

\end{document}